\documentclass[conference]{IEEEtran}
\IEEEoverridecommandlockouts
\usepackage{cite}
\usepackage{amsmath,amssymb,amsfonts}
\usepackage{algorithmic}
\usepackage{graphicx}
\usepackage{textcomp}
\usepackage{xcolor}
\usepackage{url}
\usepackage{soul}
\def\BibTeX{{\rm B\kern-.05em{\sc i\kern-.025em b}\kern-.08em
    T\kern-.1667em\lower.7ex\hbox{E}\kern-.125emX}}

\begin{document}

\title{Framework for a Decentralized Web}

\author{\IEEEauthorblockN{Raman Singh}
\IEEEauthorblockA{\textit{School of Comp Sci \& Stats} \\
\textit{Trinity College Dublin} \\
Dublin, Ireland \\
\textit{Thapar Institute of Engineering \& Technology}\\
Patiala, India \\
rasingh@tcd.ie (raman.singh@thapar.edu)}
\and
\IEEEauthorblockN{Andrew Donegan}
\IEEEauthorblockA{\textit{School of Comp Sci \& Stats} \\
\textit{Trinity College Dublin}\\
Dublin, Ireland \\
donegaan@tcd.ie}
\and
\IEEEauthorblockN{Hitesh Tewari}
\IEEEauthorblockA{\textit{School of Comp Sci \& Stats} \\
\textit{Trinity College Dublin}\\
Dublin, Ireland \\
htewari@tcd.ie}
}

\maketitle

\begin{abstract}
Over the past decade, we have witnessed the Internet becoming increasingly centralized in the hands of a small number of giant technology firms, that control many of the most popular applications and the content they host on their platforms. In addition, in the majority of instances today, access to the Internet is usually provided through local internet service providers (ISPs) in each country. Governments in different jurisdictions can exert pressure on these technology firms and ISPs to enforce restrictions on Internet usage by their citizens, such as the blocking access to certain sites and/or content. In this paper, we present a promising new approach to circumventing some of these issues. Our decentralized web (DWeb) proposal makes use of a mesh network to connect community based routers. In addition, objects on the DWeb are indexed using blockchain technology, which allows for secure storage of immutable object references, and integrity checking of the data being served to users. Our DWeb design is also capable of operating during network partitions, and is able to quickly re-synchronize with the larger network once connectivity has been restored. The proposed concept is simulated using ns-3, Multichain and Docker technologies. The performance of the system is analysed for different cache replacement strategies with a varying number of gateway nodes. The Popularity-based cache replacement algorithm outperforms the other studied strategies, which show it is the best-suited strategy for the DWeb architecture. In summary, DWeb solves issues like data integrity, centralization and controlled access to the Internet, and the overall framework exhibits fair performance in the studied simulated environment.
\end{abstract}

\begin{IEEEkeywords}
Decentralized Internet, Blockchain, DLT, NDN, Net Neutrality
\end{IEEEkeywords}

\section{Rationale for a Decentralized Web}
Access to the Internet on a 24/7 basis is an essential component of personal, work related and leisure activity for many people around the world today. However, the current centralised nature of the world wide web can lead to impaired access in certain circumstances. Governments can censor the web, and also use the web as a surveillance tool. Increasingly, we are seeing governments around the world periodically shutting down or providing partial access to the Internet, in order to prevent the free ﬂow of information. Limiting or completely blocking access to data networks, is one tool governments can use to control both citizens and the narrative around an event.

Between July 1, 2015 through June 30, 2016 there were 81 Internet disruptions in 19 countries worldwide. These included 22 in India and Iraq, 8 in parts of Syria, 6 in Pakistan, 3 in Turkey, and 2 each in Bangladesh, Brazil, North Korea, amongst other places. Such Internet shutdowns cost at least US\$2.4 billion in GDP globally \cite{west2016internet}. A total Internet shutdown was introduced in Iran as an attempt to suppress fuel protests in November 2019 \cite{netblocks-iran-graph}. Iranian Internet connectivity went as low as 5\% during this period of time. Information could not be sent to the outside world from Iran, and the shutdown made it diﬃcult to monitor human rights violations within the country. Having a centralised Internet infrastructure that uses internet service providers (ISPs) to provide access, gives governments the power to restrict Internet activity in their jurisdictions.

A second and more imperative reason for pursuing a decentralized web is to promote \textit{net neutrality}. The net neutrality concept ensures ISPs treat all Internet communication equally without any discrimination or tiered charges. It has been observed that ISPs have processed Internet traffic differently based on parameters like content types, platform, applications, devices/communication type, source/destination of messages, and protocol \cite{gilroy-report}. In 2017, the Federal Communication Commission (FCC), USA voted in favor of repealing its own laws on net neutrality. Since then the FCC has openly opposed the idea of net neutrality and started neutralizing previous laws which promote an open Internet \cite{jesse_law_report}.

This paper presents a new design for a decentralised web (DWeb) that has no central control or point of failure. It is designed to use mesh networks to provide peer-to-peer (P2P) connections through DWeb routers, and a decentralized blockchain for securely storing immutable object references. The motivation behind this research is to create a protocol that prevents physical attacks and shutdowns of the network infrastructure, by providing more physical and logical security to the network.

The rest of the paper is organized as follows: Section II confers the related literature about decentralized networking architecture. Section III discusses the proposed architecture of the DWeb concept. Section IV, V and VI respectively explore the DWeb blockchain, new node joining and trust issues of the proposed architecture. Section VII presents the implementation detail of the DWeb concept and also analyses the obtained results. The significance of the proposed system is discussed in section VIII whereas section IX debates the future possibilities of the DWeb concept. Finally, Section X concludes the study.

\section{Related Work}
A decentralized web makes censorship more diﬃcult as there are no single points of failure such as ISPs, domain name system (DNS) servers etc., that can be easily targeted by individuals or state agencies. The decentralized web also promotes net neutrality because there are no authoritative restrictions on content by any one, apart from the original publisher of the content.

There is already a large amount of research in the public domain in the area of mesh networks combined with the blockchain technology to realize such decentralized networks. Two examples of such proposal are SmartMesh \cite{smartmesh_whitepaper2017} and RightMesh \cite{rightmesh_whitepaper2019}. The proposed networks provide connectivity in areas of limited infrastructure, where the blockchain is used for buying or selling network access. These systems provide users with Internet access without the need for external infrastructure. Unfortunately, unlike DWeb, they still provide a single point of failure, as the mobile applications providing the access to the network can be shut down by governments, and therefore restrict user access. These mesh networks differ from DWeb in the way the blockchain is being used. DWeb uses a blockchain to provide transactional security whereas in these mesh networks it is used for managing network service usage and pricing.

Named data networking (NDN) is one of five projects funded by the U.S. National Science Foundation under its "Future Internet Architecture Program". NDN has its roots in an earlier project, content-centric networking (CCN), which Van Jacobson first publicly presented in 2006 \cite{NDN-2014}. NDN seeks to create a new network layer to replace the popular internet protocol (IP). The protocol was originally designed for conversations between communications endpoints, but is overwhelmingly used for content distribution. NDN aims to remove IP datagrams containing address endpoints, and replaces them with a more general structure that can \textit{name} chunks of data in a hierarchical manner \cite{NDN-2010}. NDN is based on named content, which has no notion of a host at its lowest level, and no source and destination addresses. Because of this key feature, each content packet has a unique name, and is forwarded by a lookup of its name \cite{cdn-ndn-comparison}. Using NDN, a consumer can request content using \textit{interest packets} containing the name of the desired data. The user then receives content through \textit{data packets} containing the name of the content and the actual content itself. 

The NDN project aims to be compatible with today’s Internet infrastructure, and it is designed in such a way that IP can run over NDN and vice versa \cite{NDN-2010}. This inclusion of NDN causes large overheads. In NDN, the maximum payload of 4096 is allowed in default packets and includes a 550 byte header and the interest packet segment with sizes from 150 to 250 bytes. Thus, transmitting a NDN packet needs four Ethernet frames, causing the total overhead to be about 23.6\%. The overhead is caused by the re-transmission of lost packets which is more expensive in NDN, due to the large 4096 bytes chunk payload \cite{cdn-ndn-comparison}. In NDN, names are longer and more numerous than IP addresses, which makes NDN routing tables much larger, which in turn results in inefficient lookups \cite{cdn-ndn-comparison}. In this networking approach, blockchain is not used for any purpose, whereas DWeb uses a blockchain for providing sdata integrity.

Skywire \cite{skywire-project} is a promising project which challenges the centralized Internet by creating a community driven mesh network. In this project, users own their own infrastructure which enables them to enjoy fast, reliable and private Internet access. The network is powered by various nodes which are maintained by individual users. In this type of system, each individual station will have hardware which can act as router also. The hardware is used to access the information and can act as an intermediate node to forward information to nearby nodes. This setup gives greater power in the user but at the same time makes routing complex. The number of routing nodes increases manifold if all users participate in routing and storing blockchain data. DWeb provides a similar level of freedom to its users but lowers the routing overheads from the personal device of its users. DWeb focuses on community-owned routing devices and hence offers a robust and flexible routing option at the same time.

The Althea project \cite{althea-project} allows users to set their mobile nodes (e.g. personal smartphones) as decentralized and manageable community ISPs. These pseudo-ISPs can be allowed to be used as a community router in the networked ecosystem. The basic technology used in this project includes routing protocol (known as firmware) and hardware infrastructure. These routers can automate network configuration and pay other routers for their bandwidth. These community routers are connected to each other and serve as a collective network. The Althea community network is then connected with commercial Internet which serves as gateway to the world wide web. This project does not use the potential of blockchain to solve the security issue one can face due to community owned routing devices. There must be some level of trust among communities to route personal data on these routing devices and DWeb solves this problem efficiently.

BlockMesh \cite{blockmesh-project} is an anonymous P2P, decentralized communication platform. BlockMesh is an integration of blockchain technology and mesh networking primarily developed for Internet of Things (IoT) devices. The BlockMesh framework is used to share files, send text messages and digital transactions, and for making voice calls. Users with unlimited Internet access can share their bandwidth with others who do not have any Internet access, using mesh technology. DWeb provides many more services than this P2P platform as our proposed system does not believe in owning any Internet data.

\section{System Architecture}
\begin{figure}[htbp]
\centering
\includegraphics[width=3.5in]{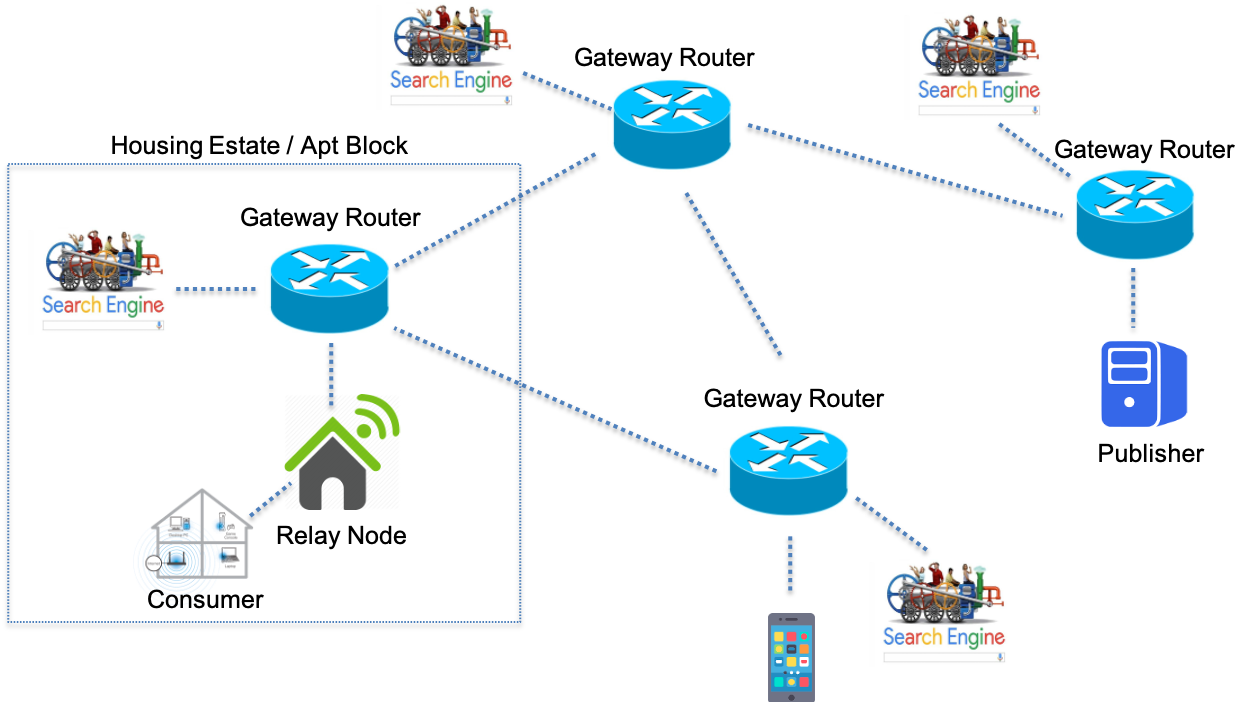}
\caption{DWeb System Architecture}
\label{fig:dweb-sys_arch}
\end{figure}

Blockchain technology has shown how we can move from a centralized decision making network to a distributed, transparent and more inclusive one. In our DWeb protocol we wish to apply the same philosophy and create a decentralized web infrastructure for static objects. Our system makes use of mesh networks and point-to-point protocol (PPP) links, which aims to prevent physical attacks to the network, and creates highly replicated and immutable blockchain backed data stores. Such data stores prevent logical attacks to access to information. A high-level view of the system architecture can be found in Figure \ref{fig:dweb-sys_arch}.

\subsection{Network Topology}
A typical DWeb network topology may, for example, consist of a housing estate or an apartment block, where each household has their own DWeb router which wirelessly connects to a neighbouring DWeb router, which in turn connects to another router etc. One or more routers in the estate will act as the \textit{Gateway Router}. Such routers may have a more powerful ``directed wireless antenna" to one or more other DWeb gateway routers in the adjoining neighborhood. In this way we are able to form a mesh network of DWeb gateways, clustered around which there will be a number of users. Search engines in the system design are publicly accessible entities attached to DWeb gateways, to allow for efficient access to objects. A consumer can query a \textit{Search Engine} to look for objects on the DWeb. A publisher will advertise its objects via its nearest DWeb gateway, which can then satisfy requests from consumers for that object.

\subsection{Protocol Stack}
\begin{figure}[htbp]
\centering
\includegraphics[width=1in]{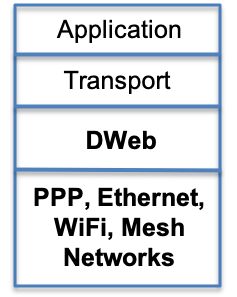}
\caption{The DWeb Protocol Stack}
\label{fig:dweb-net_stack}
\end{figure}

DWeb routers run a protocol stack as shown in Figure \ref{fig:dweb-net_stack}. The stack consists of a new DWeb network layer which replaces the existing IP layer, while the functionality of the upper layers (Application and Transport) remains as before. The DWeb protocol makes use of medium access control (MAC) addresses and has a \textit{flat address} space, as opposed to the hierarchical address model that is used in IP. Communication links operate on a PPP basis using available wired and wireless technologies.

\subsection{Addressing and Routing}
Each DWeb router is connected to at least one other DWeb router in the system. When each router follows this rule, a mesh network is created. The DWeb protocol uses MAC addresses to replace IP to route packets via PPP links. Mesh networks interconnect over IEEE 802.11s. This creates a wireless local area network (WLAN) network \cite{IEEE-802-11}. 802.11s is based on the 802.11 medium access control (MAC) protocol. The 802.11 protocol uses link layer (i.e. MAC) addresses as source and destination addresses.

Creating a local network using mesh networks and routing via MAC addresses, removes the need for hierarchical IP addresses and therefore DNS lookups. This in turn eliminates the need for DNS servers to be hosted by ISPs. DNS servers can be a single point of attack in a controlled shutdown of the Internet. With mesh networks implementing a PPP protocol and sending data directly from one router to another, there is no requirement for ISPs to route the traffic. Removing ISPs from the equation diminishes the ability of governments to perform Internet shutdowns by instructing ISPs to stop providing their service. DWeb gateways also maintain paths to other popular gateways in the DWeb, by making use of a time and space limited cache. Gateway routers maintain a synchronized copy of the blockchain ledger which contains a fingerprint of each object that has been uploaded to the DWeb.

Finally, each DWeb gateway has a publicly accessible search engine attached to it which consists of transactions extracted from the synchronised ledger, for quicker access to objects on the DWeb. There are two types of edge nodes in the system: \textit{publishers} and \textit{consumers}. A publisher is an edge node that publishes web objects. A publisher sends a copy of any new object to its default DWeb gateway to be included in the blockchain. Publishers may have a direct or indirect connection to a DWeb router. A consumer has knowledge of gateways on the DWeb, and can query search engines to locate objects and requests its default DWeb gateway to retrieve objects.

\section{The DWeb Blockchain}
A blockchain is used in the system for immutable data indexing of objects. The DWeb blockchain is a \textit{permissioned} ledger to which blocks can only be added by authorized DWeb gateways. The opportunity to add a new block is controlled in a round robin fashion, thereby eliminating the need to perform a computationally intensive proof-of-work (PoW) puzzle.

\begin{figure}[htbp]
\centering
\includegraphics[width=2in]{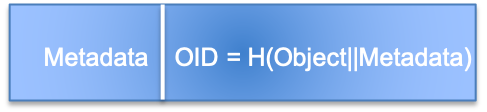}
\caption{Blockchain Transaction}
\label{fig:blockchain_trans}
\end{figure}

When a new object is created by a publisher, it sends a copy of the object and its associated metadata to it's default gateway router. The router then creates a new blockchain transaction and broadcasts it to the P2P network. A blockchain transaction consists of the object's \textit{metadata} and a \textit{hash} of the concatenation of the ``object" and its ``metadata", as shown in Figure \ref{fig:blockchain_trans}.

The cryptographic hash acts as an integrity check, as well as a unique object identifier (OID), and represents a \textit{flat} name space. Each new transaction must be digitally signed with the secret-key of the DWeb gateway that originates it. The signature can be verified by all other DWeb gateways, by accessing the public-key certificate of the router that originated the transaction. Unsigned transactions are automatically rejected by the network, and this prevents content pollution by unauthorized publishers.

\section{Joining the DWeb}
A new gateway router joining the DWeb must download a full copy of the blockchain. It indexes transactions from the blockchain into its publicly accessible search engine according to a known algorithm, so as to mirror other DWeb search engines. Every new gateway router also obtains a public-key certificate from a commercial certification authority (CA) \cite{ca}. The certificate binds the MAC address of a DWeb gateway to its public key and prevents MAC address spoofing by malicious nodes. The certificate is broadcast to the P2P network and locked into the blockchain as a transaction. It can be accessed by all other DWeb gateway routers, who can use it to verify that a transaction did in fact originate from an authorized DWeb gateway router.

\subsection{Adding a New Object}
\begin{figure}[htbp]
\centering
\includegraphics[width=3.5in]{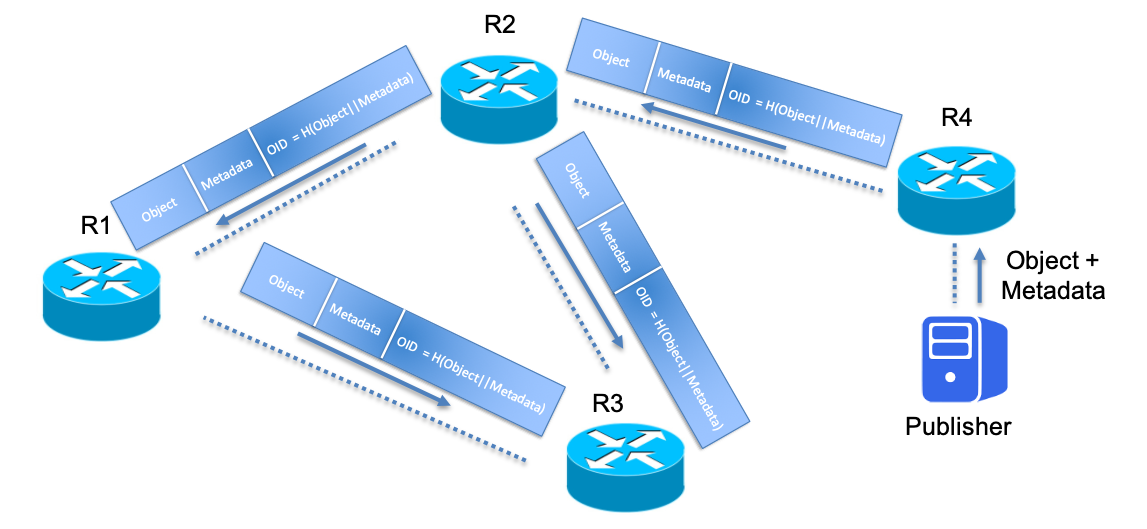}
\caption{Adding a New Object to the DWeb}
\label{fig:new_trans}
\end{figure}

On receiving a new object from a publisher, a DWeb gateway creates and broadcasts a new blockchain transaction consisting of the OID and Metadata, along with the object itself to the P2P network, for the transaction to be included in the next block as seen in Figure \ref{fig:new_trans}. Once a transaction has been verified by other nodes it is locked into the blockchain. Other gateways can cache the address of the \textit{upstream router} from where the transaction originated. Gateway routers can add an entry for the object into their search engine, and \textit{optionally} store a copy of the object in their local databases.

\subsection{Object Retrieval}
The easiest way to find an object is for a consumer to query a DWeb search engine. The engine returns a set of results (i.e. links) based on the search query, where each result consists of an object identifier and associated metadata:

\begin{center}
${Link = (OID, Metadata)}$  
\end{center}

\begin{figure}[htbp]
\centering
\includegraphics[width=3.5in]{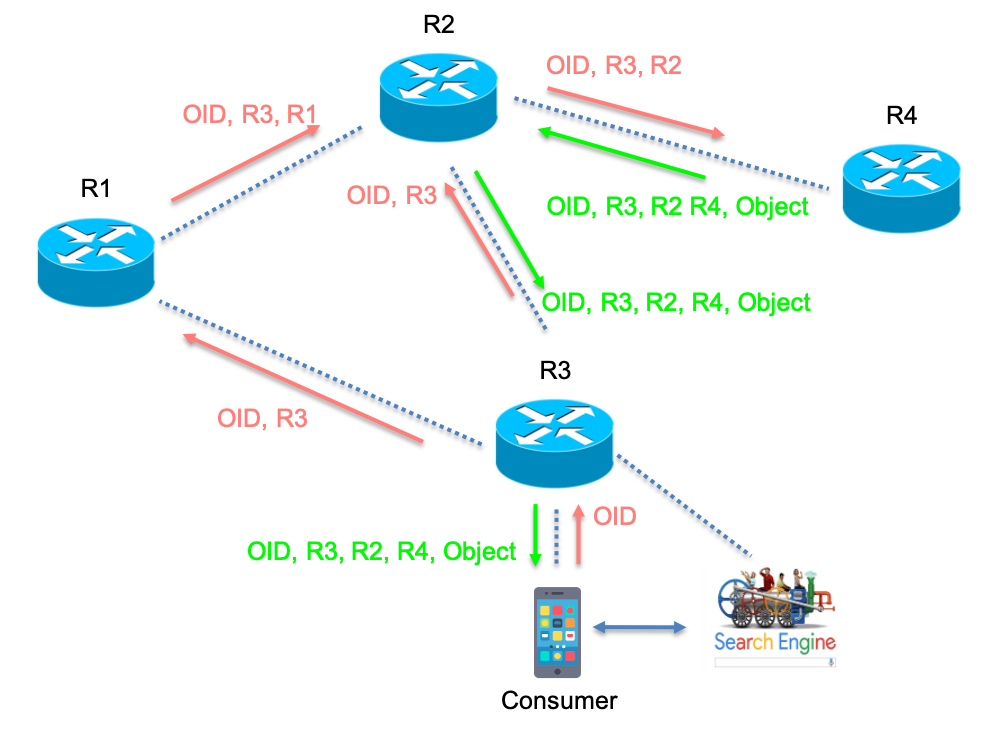}
\caption{Fetching a Non-Cached Object}
\label{fig:fetch_object}
\end{figure}

On clicking a \textit{Link}, a request will be sent out to the default router (possibly via one or more relay nodes) which will retrieve the object. If an object is cached in the default router's search engine, the object can be served immediately. If not, a request will be broadcast on the P2P network with the object's OID and router's address. Each router along the path adds it's address prior to forwarding the request. A router that has the specified object can satisfy the request by simply reversing the path to get the object to the requesting consumer, as shown in Figure \ref{fig:fetch_object}. This is similar to the dynamic source routing (DSR) \cite{dsr-protocol} protocol. However, the addresses used in this projects design are MAC addresses rather than IP addresses.

In Figure \ref{fig:fetch_object}, a consumer requests an object from the DWeb. The request is broadcast to the P2P network as represented by the \textit{red} arrows. In this case the request is satisfied by router R4 and the object is sent back by reversing the request path, as represented by the \textit{green} arrows. The consumer node is able to compare the OID of the retrieved object with the OID it was given by the search engine. It concatenates the fetched object and with the metadata which is contained in the \textit{Link} and passes it through a hash algorithm. If the resulting hash matches the OID in the \textit{Link} then the consumer can be confident that they have been served the correct object, and that the object has not been modified in any way.

\subsection{Network Partition}
\begin{figure}[htbp]
\centering
\includegraphics[width=3.5in]{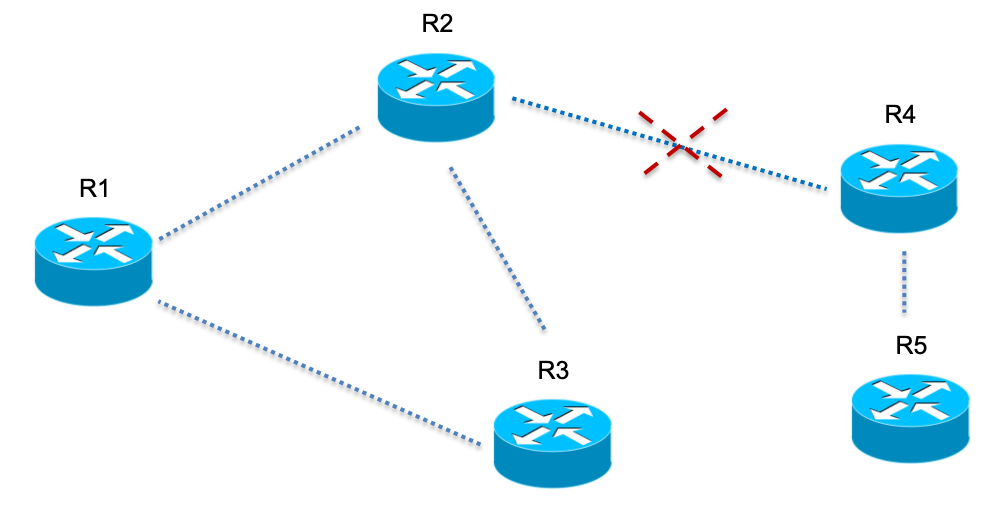}
\caption{Network Partition}
\label{fig:net_part}
\end{figure}

It is possible that parts of the network may lose connectivity to the DWeb from time to time. A node may rejoin the network at a later stage or never reconnect with the network. During a network partition all updates to the ledger will be ``local” to the routers in the sub-network. When connectivity is re-established with the DWeb the sub-network nodes must use the \textit{longest blockchain} being advertised. The longest blockchain is determined by the number of object storage transactions that have taken place, not the number of nodes in the network. The reconnected nodes broadcast any ``local transactions” to the P2P network for them to be included in the longest chain. In Figure \ref{fig:net_part}, routers R4 and R5 become disconnected from the main network.

All updates to the ledger in routers R4 and R5 will be local to their sub-network. When connectivity is re-established, the longest blockchain is used. In this example this will be R1, R2 and R3's chain. Therefore the ``local” transactions from R4 and R5 will be advertised to be included in the longer chain. Each router in the network will then have a synchronised copy of the blockchain with all transactions that happened during the partition.

\section{Trust}
DWeb gateway routers are \textit{trusted} entities in the system. Other nodes in the network rely on gateway nodes to operate in a fair manner, by creating blockchain transactions on their behalf, and correctly populating their search engines. This trusted privilege is granted to gateway routers once they obtain a public-key certificate form a trusted CA, and register the certificate as a transaction on the blockchain.

However, this privilege can also be rescinded by the collective DWeb community, by revoking the public-key certificate of a ``blacklisted" gateway, if for example the \textit{trust score} of the router falls below a certain threshold. Certificate revocation can be easily achieved on the DWeb by adding a ``dummy" certificate into the blockchain for the MAC address associated with the erring router. This responsibility could be assigned to one or more ``super nodes" in the network. All nodes on the network will then use the \textit{new} public key certificate to verify any new transactions that originate from the blacklisted router. Since the blacklisted router will no longer have the correct corresponding secret key, it will not be able to correctly sign any new transactions, in order for them to be included in the blockchain. 

All other routers on the DWeb also have the ability to become a gateway router by obtaining a public-key certificate of their own from a trusted CA. Therefore, a publisher can run their own gateway router and forward to it any object references that it wishes to be added to the blockchain. Similarly, a consumer can run a full gateway node and keep a synchronized copy of the blockchain from which it can search for objects on the DWeb. In this manner, a consumer or a publisher no longer have to trust other network entities, while retrieving or adding objects to the DWeb.

\section{Implementation and Results}
\begin{figure}[htbp]
\centering
\includegraphics[width=3.2in]{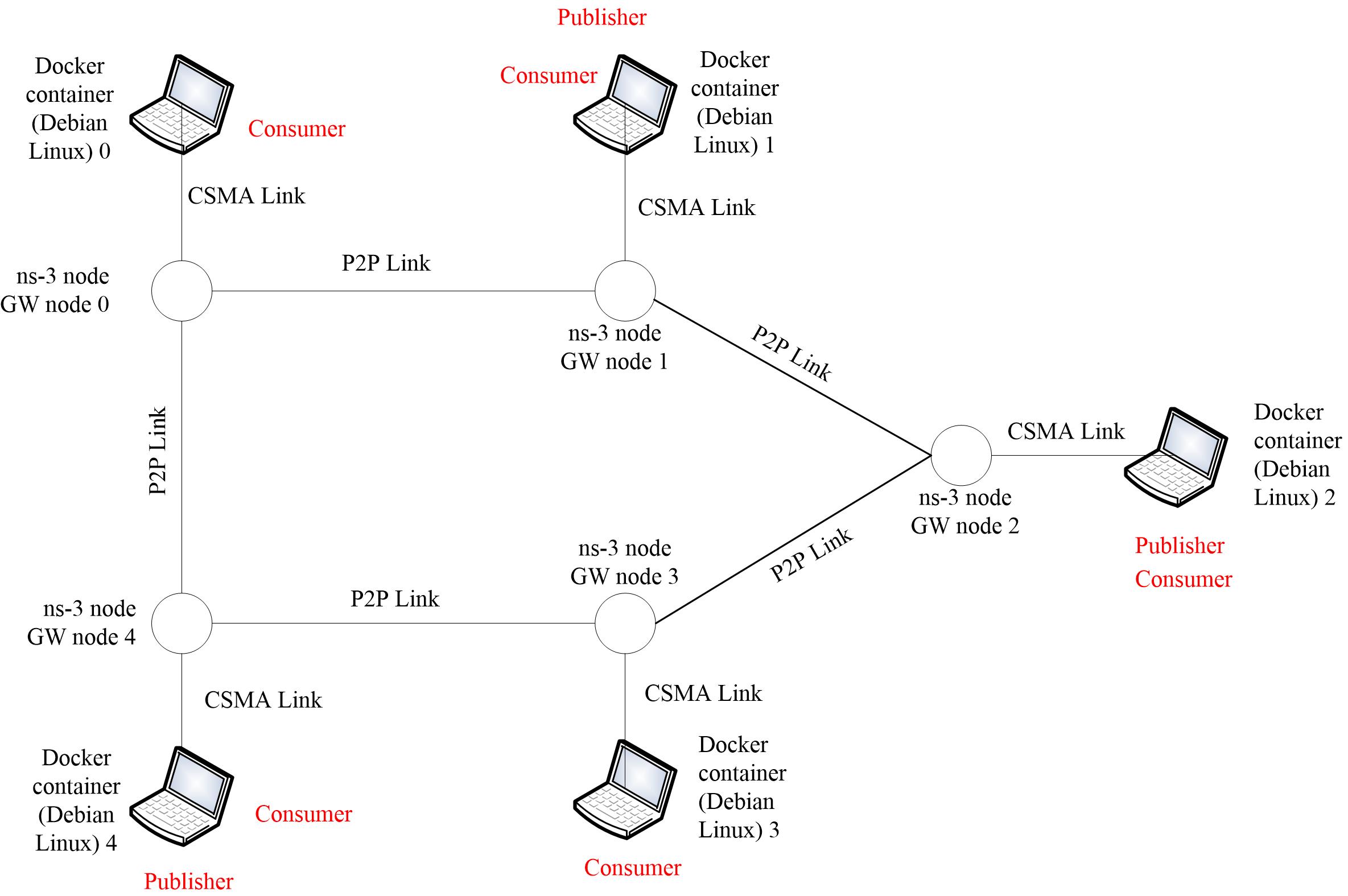}
\caption{DWeb Simulation Architecture}
\label{fig:imp_archi}
\end{figure}

Our proposed DWeb architecture has been simulated using the ns-3 \cite{nsnam-ns3}, Multichain \cite{multichain-coin} and Docker \cite{docker} platforms. The basic architecture for the simulation environment is shown in Figure \ref{fig:imp_archi}. The gateway nodes (which act as routing nodes) are actually ns-3 nodes which are connected via P2P links. ns-3 does not provide native blockchain functionality, and in order to utilize it each gateway (GW) node is connected to a Docker container which runs Debian Linux.

The various entities/functionality of the DWeb such as producers, consumers, object handling, trust and object integrity are implemented within the Docker environment using the Python programming language. A producer calculates the hash of an object along with its metadata. This hash acts as an OID and is stored on the blockchain along with metadata of the object. The object is then published on the associated GW node and stored in its cache. When a consumer searches for an object for the first time, the original producer acknowledges the \textit{interest} because only that particular GW node holds it in its cache. Repeated interest requests of any object by other consumers will result in the fetching of the object from either original producer or intermediate GW nodes, depending on where it is located and the distance between them. When the GW node acknowledges the presence of a requested object in its cache, the consumer verifies the integrity of object by calculating hash of object plus the metadata, and comparing it with the hash stored in the blockchain (known as object verification), to ensure the integrity of the data. 

\begin{figure}[htbp]
\centering
\includegraphics[width=3.8in]{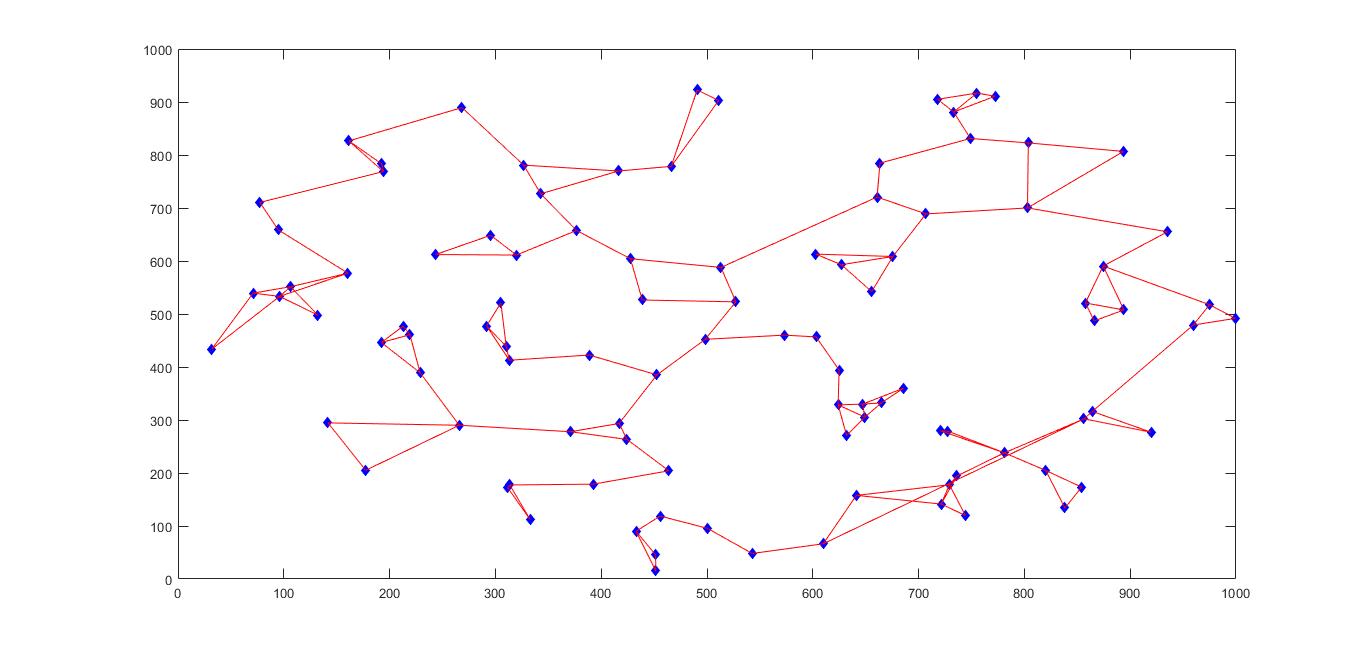}
\caption{Topology for 100 Gateway Nodes}
\label{fig:topo_100}
\end{figure}

The DWeb concept has been tested using a scenario with 10, 50 and 100 GW nodes. The topology of the 100 GW node scenario is shown in Figure \ref{fig:topo_100}. The scenario assumes that one GW node is situated in each locality, and wireless access points of that locality are directly connected to that particular node. Four different cache replacement strategies i.e. Popularity-based, First in First Out (FIFO), Least Recently Used (LRU), Least Frequently Used (LFU) have been implemented to analyse the performance of the DWeb system. For each of the experiments the cache size for all replacement policies is set to 1 GB.

\begin{figure}[htbp]
\centering
\includegraphics[width=3.3in]{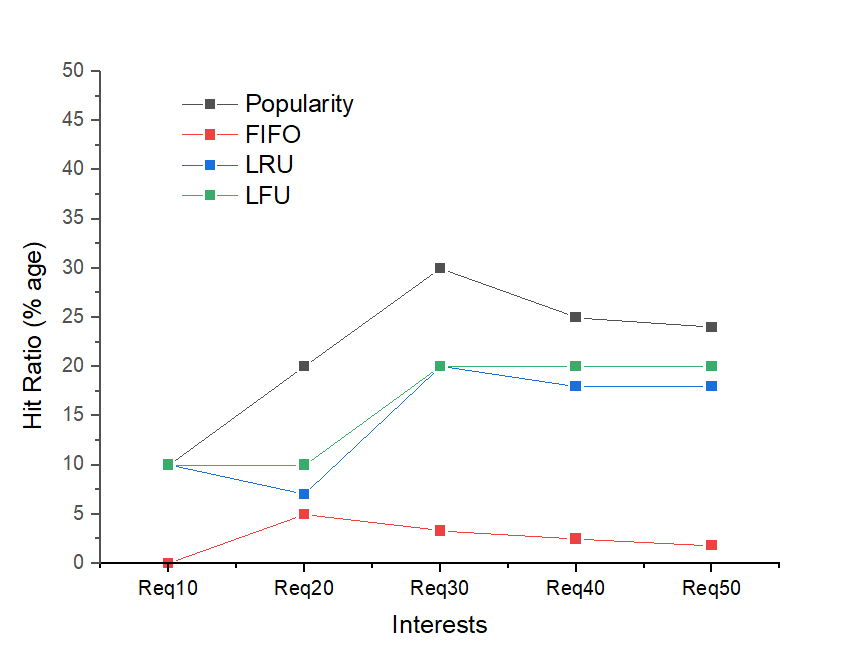}
\caption{Comparative Analysis of Various Cache Replacement Strategies}
\label{fig:comp_cache}
\end{figure}

Figure \ref{fig:comp_cache} shows the comparative analysis of various cache replacement strategies for DWeb. The x-axis shows the number of requests each gateway node made for objects, whereas the y-axis depicts the hit ratio obtained by particular cache replacement algorithm.  In the Popularity-based cache replacement strategy, when a consumer ``enters the interest in the DWeb" (i.e. searches for an object), and if the object is popular, the object is cached not only in the consumer node, but also in cache of intermediate GW nodes along the path. Because of this preemptive caching of popular objects, the Popularity-based cache replacement strategy outperforms the other algorithms like FIFO, LRU and LFU.

\begin{figure}[htbp]
\centering
\includegraphics[width=3.3in]{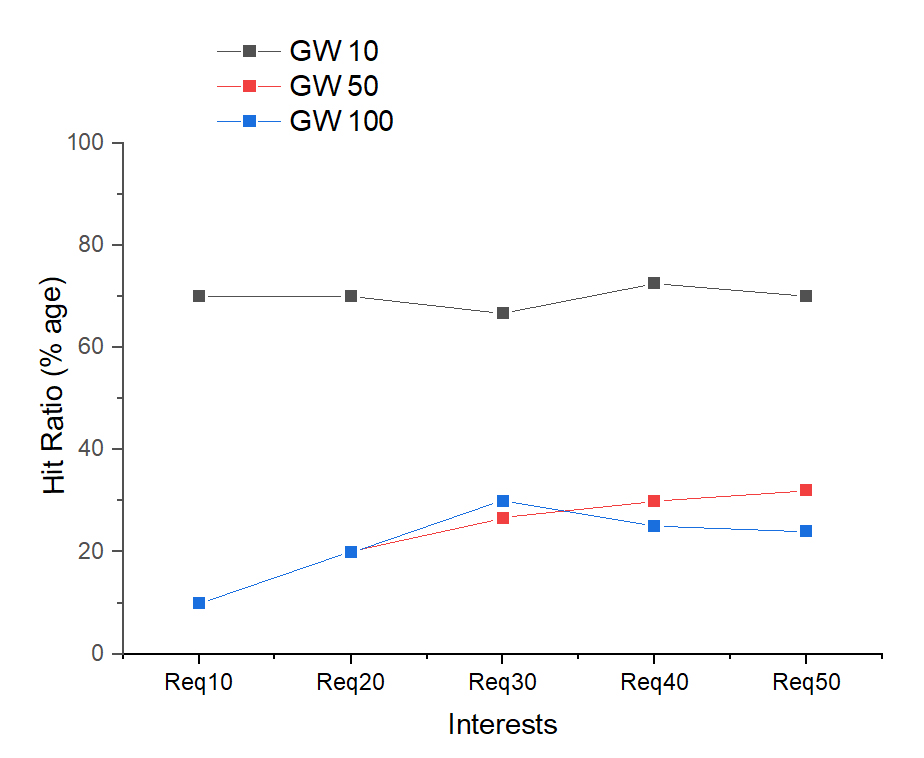}
\caption{Hit Ratio for Popularity-based Caching for 10, 50 \& 100 GW Nodes}
\label{fig:hit_gw}
\end{figure}

Amongst the worst performing cache replacement strategies is the FIFO strategy. One reason behind this is that this algorithm replaces the first cache entry even if that object is the most popular object in the cache. LRU and LFU offer similar performance, however it is found that LFU gives a slightly higher hit ratio than LFU. These results are obtained if the popularity parameter is fixed at 60\%, i.e. during the simulation 60\% of all generated requests are for popular objects. Figure \ref{fig:hit_gw} shows the variation of the hit ratio for the Popularity-based caching algorithm when the number of requests are varied for different numbers of GW nodes. The results show that the hit ratio is stable for fewer number of GW nodes, whereas it increases as the number of requests are ramped up for GW nodes 50 and 100.  

In the simulation a smaller number of GW nodes means that fewer number of requests are generated, which increases the hit ratio, as the repeating interests are high. This is why the overall hit ratio is higher for the 10 GW node experiment, as opposed to other experiments with higher numbers of GW nodes. 

\section{DWeb Distinctions}
To the best of our knowledge the DWeb is the only decentralized web architecture which uses blockchain technology. This functionality is used for object verification which ensures the data integrity of the objects accessed by consumers. This capability can also be used to discard the old version of objects, and force GW nodes to store updated versions of objects. The nearest network architecture to DWeb is NDN but both have differences in how objects are accessed, data integrity is verified etc. The DWeb can handle network partitions more efficiently as blockchain is coherent and can work even if link failures force it to fork temporarily. Different partitions of network topology can work independently in the DWeb, and when the link is restored all records are seamlessly updated in all GW nodes. In DWeb, security is an intrinsic part of its architecture as it provides authentication, authorization and integrity of the transactions, which is missing from any other Content Centric Networks (CCN). Moreover, DWeb has the ability to discard unauthorized publishers using public-key certificates which protects the network from content poisoning. Unlike all other CCNs, only DWeb has the capability to inherently authenticate the joining of a new GW node, which can safeguard network from various attacks like man-in-the-middle, distributed denial of service, node capture or node impersonation attacks.  

\section{Future Work}
The initial framework of DWeb is proposed in this paper, but there are still some core research problems that need to be addressed in order to make this blueprint exhaustive. These meaningful challenges include mobile consumers and publishers, object storage, cache limitations, efficient routing and superlative implementation.

The \textit{trust score} can be used to find unreliable or biased routers on the DWeb. Such routers should score less than their counterparts, and their certificate can be revoked as mentioned earlier. Efficient calculation of the \textit{trust score} is an open-ended research problem and needs comprehensive investigation.

In order to investigate the proposed DWeb architecture, we have tested the basic concepts using the ns-3, Multichain and Docker platforms. In this implementation, some of the links are IP-based, e.g. GW to Docker container links. We are now designing and implementing a simulation test bed using the ndnSIM \cite{ndnSIM}, Ethereum \cite{ethereum} and Docker platforms, which will completely remove the IP-based system. Comparative real-time scenarios using ns-3 nodes resembling physical routers will be developed in order to validate the DWeb architecture.

\section{Conclusion}
The Internet as we know it today is controlled by technology behemoths and ISPs. The primary motivation of these organizations is to generate profits for their shareholders, and in some cases they are beholden to the whims of the local government agencies. The decentralized web (DWeb) is proposed to circumvent some of the shortcomings of a centralized Internet architecture, such as censorship of content or denial-of-service. The DWeb uses the blockchain technology to store immutable references to objects created by publishers, which are shared by routers connected by mesh and point-to-point protocol links. The routers are community owned and no one entity directly controls their functionality. This ensures for the neutral operation of the DWeb.

A consumer can efficiently search for objects by using publicly available search engines on the DWeb. In the event of network link failure, unlike the centralized web, all of the disassociated networks continue to share objects. Once the link is restored, the DWeb can update its entire ledger through an effective ledger regeneration mechanism. The DWeb has potential to alleviate the challenges caused by today's centralized web, and has the power to revolutionize the way people use and perceive the Internet in the future.

\end{document}